# SuperDTI: Ultrafast diffusion tensor imaging and fiber tractography with deep learning

Hongyu Li[1], Zifei Liang[2], Chaoyi Zhang[1], Ruiying Liu[1], Jing Li[3], Weihong Zhang[3], Dong Liang[4], Bowen Shen[5], Xiaoliang Zhang[1], Yulin Ge[2], Jiangyang Zhang[2], Leslie Ying[1,6]

[1]Electrical Engineering, University at Buffalo, State University of New York, Buffalo, NY, USA

[2]Center for Biomedical Imaging, Radiology, New York University School of Medicine, New York, USA

[3]Radiology, Peking Union Medical College Hospital, Peking Union Medical College and Chinese Academy of Medical Sciences, Beijing, China

[4]Paul C. Lauterbur Research Center for Biomedical Imaging, Medical AI research center, SIAT, CAS, Shenzhen, P.R.China

[5]Computer Science, Virginia Tech, Blacksburg, VA, USA

[6]Biomedical Engineering, University at Buffalo, State University at New York, Buffalo, NY, USA

**Correspondence**

Leslie Ying, Ph.D.

Department of Biomedical Engineering, Department of Electrical Engineering, University at Buffalo, State University of New York

223 Davis Hall, Buffalo, NY, 14260, USA

Email: leiying@buffalo.edu

**Purpose:** To propose a deep learning-based reconstruction framework for ultrafast and robust diffusion tensor imaging and fiber tractography.

**Methods:** We propose SuperDTI to learn the nonlinear relationship between diffusion-weighted images (DWIs) and the corresponding tensor-derived quantitative maps as well as the fiber tractography. Super DTI bypasses the tensor fitting procedure, which is well known to be highly susceptible to noise and motion in DWIs. The network is trained and tested using datasets from Human Connectome Project and patients with ischemic stroke. SuperDTI is compared against the state-of-the-art methods for diffusion map reconstruction and fiber tracking.

**Results:** Using training and testing data both from the same protocol and scanner, SuperDTI is shown to generate fractional anisotropy and mean diffusivity maps, as well as fiber tractography, from as few as six raw DWIs. The method achieves a quantification error of less than 5% in all regions of interest in white matter and gray matter structures. We also demonstrate that the trained neural network is robust to noise and motion in the testing data, and the network trained using healthy volunteer data can be directly applied to stroke patient data without compromising the lesion detectability.

**Conclusion:** This paper demonstrates the feasibility of superfast diffusion tensor imaging and fiber tractography using deep learning with as few as six DWIs directly, bypassing tensor fitting. Such a significant reduction in scan time may allow the inclusion of DTI into the clinical routine for many potential applications.

## 1. INTRODUCTION

Diffusion-weighted imaging (DWI) uses diffusion sensitizing gradients to measure the extent of water molecule diffusion along the gradient direction.[1-3] DWI has shown to be useful for early detection of ischemic stroke,[1,4] as well as other brain diseases, such as multiple sclerosis,[5-7] trauma,[8,9] brain tumors,[10,11] and hypertensive encephalopathy.[12,13] To account for intricate patterns of water diffusion profiles shaped by tissue microstructural organization, e.g., anisotropic diffusion in white matter structures, diffusion tensor imaging (DTI) was later developed to characterize three-dimensional tissue water diffusion using a second-order diffusion tensor model.[14] From DTI data, several markers, such as mean diffusivity (MD) and fractional anisotropy (FA), can be derived and are widely used to visualize microstructural organizations in the brain as the diffusion tensor itself, a three-by-three matrix, is difficult to visualize. Furthermore, the 3D directional anisotropy information encoded in diffusion tensors allows non-invasive reconstruction of the trajectories of major white matter tracts in the brain.[15] Although more sophisticated diffusion MRI techniques, such as diffusion kurtosis imaging[16] and high angular resolution diffusion imaging (HARDI),[17] have been developed to provide more comprehensive information on tissue microstructure (e.g., non-Gaussian diffusion and fiber crossing), DTI remains an important tool for neuroscience research with a wide array of clinical applications.

Although DTI theoretically requires only six diffusion-weighted images and one non-diffusion-weighted image for estimation of the diffusion tensors, a large number of diffusion-weighted images with different diffusion encoding directions are often acquired in practice due to the low SNR and high sensitivity of the tensor model to noise contamination. For example, 30-90 DWIs are typically needed to obtain MDs and FAs with diagnostic quality, resulting in a scan time of 10-30 minutes. Such a prolonged scan time can increase motion artifacts and the patient's discomfort. Several techniques have been developed to accelerate DTI, such as parallel imaging-based simultaneous multislice (SMS) acquisition,[18] and compressed sensing.[19-24] However, the acceleration factor is limited to 2-3, with the latter involving extensive computation power.

Here we report a deep learning approach to shorten the acquisition time of DTI dramatically. This approach can be applied on top of SMS and requires little online computation once the training is completed offline. We used deep convolutional neural networks to model the nonlinear relationship between the acquired DWIs (noisy and might be corrupted due to k-space undersampling) and the desired DTI-derived maps, and the speed improvement was achieved by reducing the number of DWIs and the amount of k-space data of each DWI at the same time. Although machine learning has been used to accelerate diffusion MRI,[25,26] at least 12 DWIs were needed to generate DKI and NODDI-derived maps[25] and 25 DWIs for fiber tracking.[26] In Golkov et al,[25] a multi-layer perceptron (MLP) network (called q-space deep learning (q-DL) here) was used to learn various diffusion parameters from the q-space samples at each pixel without imposing diffusion models. The q-DL and other subsequent studies have demonstrated the potential of using machine learning to reduce the q-space data necessary for diffusion imaging.[27-32] Besides multi-layer perceptrons,[25,27,31] some used convolutional neural networks.[26,28-30,32-35] In Tanno et al,[35] a convolutional neural network was used to combine the intrinsic and parameter uncertainty quantify the predictive uncertainty of deep-learning super-resolution DWIs such that the diffusion parameters can be obtained with more reliability. The recent DeepDTI method used a convolutional neural network to improve the quality of DWIs.[34] After high-quality DWIs were obtained from raw DWIs, T1 and T2-weighted images, they were fitted to the conventional tensor model.[34] Although DeepDTI demonstrated the feasibility of fast DTI using deep learning, it still needs to fit the tensor model using the conventional method, which is well-known for lack of robustness to noise/errors.[36]

In this study, we develop a deep convolutional neural network to generate FA/MD, directionally encoded colormaps, and fiber tractography using as few as 6 DWIs. These DWIs can be noisy or misregistered by motion. We name our approach SuperDTI. Some of the preliminary results have been reported in conference abstracts.[29,30]

SuperDTI is different from q-DL and its derivatives in that a novel residual-learning convolutional neural network is designed to generate the quantitative maps directly from the entire DWIs (instead of pixel-wise calculation in Golkov et al[25]) such that the correlation in the image domain and q-space are both utilized. SuperDTI is also different from DeepDTI and Tanno's method in that it completely avoids the error-sensitive tensor fitting step.[36] With experimental results, we demonstrate the feasibility of highly accelerated (15 × acceleration) DTI and fiber tracking using our proposed SuperDTI method, bypassing all models. Our results show that as few as 6 raw (without pre-processing like denoising) DWIs are needed for FA maps and fiber tractography, and 3 raw DWIs for MD maps. Experiment results also show that the proposed method is much less sensitive to noise than the conventional tensor fitting method and is robust to misregistration due to motion among DWIs. Furthermore, we verify that networks trained using healthy subjects can potentially be applied to stroke patients with lesions.

## 2. THEORY

### 2.1 Network architecture

Our objective is to represent the nonlinear mapping between q DWIs (input) and the FA, MD, or FA color maps (output) using a deep neural network, bypassing the tensor fitting process. If the nonlinear mapping between the input x and the output y is represented as $y = F(x; \Theta)$, where $\Theta$ is the parameter that controls the nonlinear relationship, then a deep neural network is designed such that the parameter $\Theta$ can be learned to represent the true relationship through training. Figure 1 shows the basic architecture of the convolutional neural network (CNN) used in SuperDTI. In the deep learning method, CNN uses an input layer, an output layer, and multiple hidden layers of nodes to form a hierarchical structure. Each node of a layer is connected to some nodes of the previous layer by a linear convolution, a nonlinear activation, or a pooling (reduction from multiple to one) process. Such a hierarchical structure with deep layers can represent highly complex nonlinear models, with different network parameters representing different models.

SuperDTI is a deep U-net[37] style encoder-decoder network with residual learning[38] and patch-wise training/padding.[39,40] The skip connection, originally introduced in ResNet,[38] copies the feature maps from early convolutional layers and reuses them as the input to later deconvolutional layers of the same size in a network. There are four conveying paths in our networks that enable residual learning to boost CNN performance.[38,39,41,42] Except for the last layer, each convolutional or deconvolutional layer is followed by a rectified linear unit (ReLU).

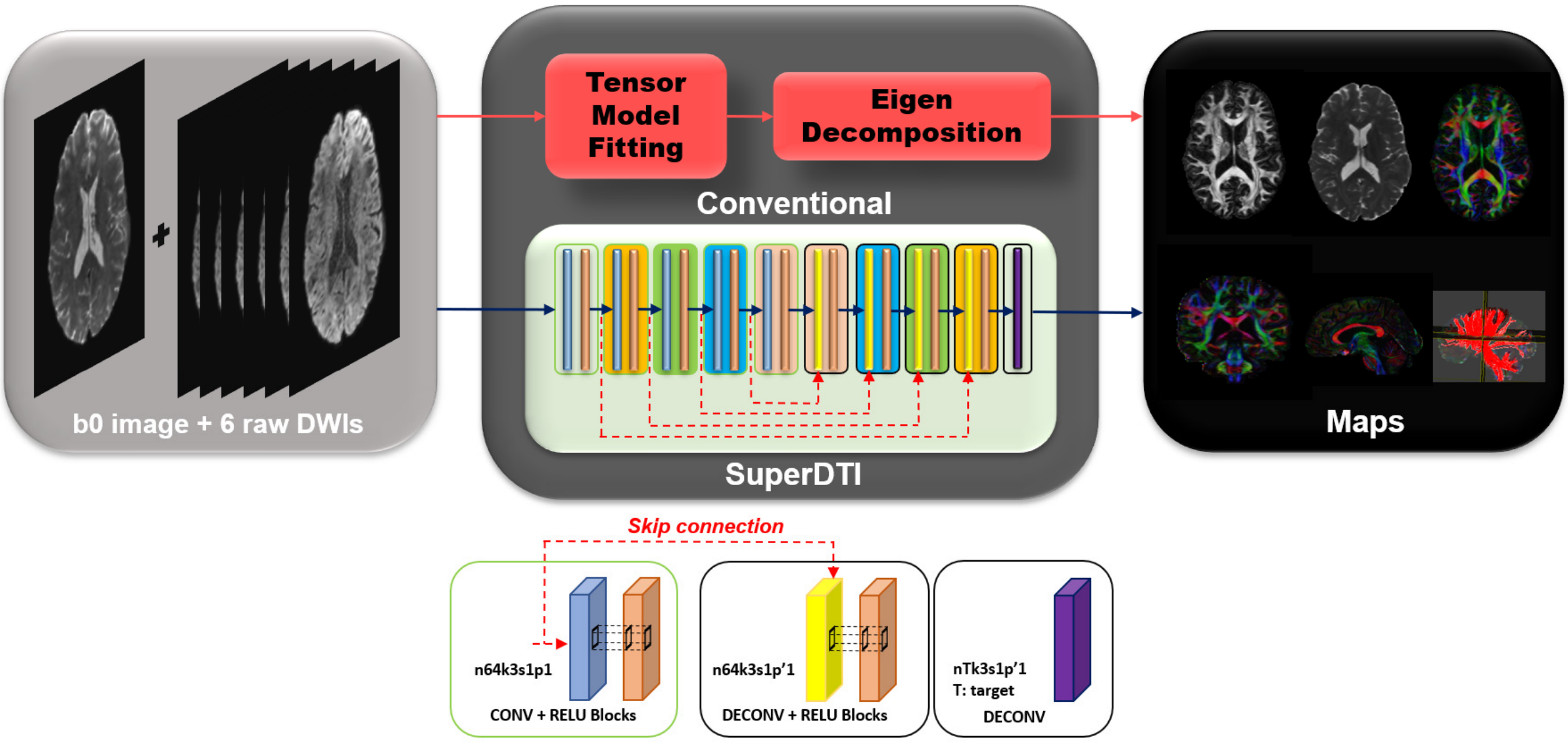

**FIGURE 1** Schematic comparison of the conventional DTI model fitting and deep learning methods SuperDTI for generating various diffusion quantification maps. The proposed network comprises several layers of a skip-connection-based convolution-deconvolution network which learns the residual between its input and output. In each layer, $n64k3s1p1$ (p'1) indicates 64 filters of kernel size 3 × 3 with a stride of 1 and padding of 1 (truncation of 1). Except for the last layer, each (de)convolutional layer is followed by a ReLU unit.

We use a $L$ layers (de)convolutional neural network architecture. Specifically, each of the $l = 1, \ldots, L$ hidden layers calculates

$$H_l = \sigma_l(W_l * H_{l-1} + B_l) \qquad (1)$$

for layers without skip connection, and

$$H_l = \sigma_l(W_l * (H_{l-1} + H_{L-l}) + B_l) \qquad (2)$$

for layers with skip connection, where $H_l$ is the output of layer $l$ ($H_0$ stands for input), $W_l$ and $B_l$ represent the filters and biases respectively, '*' denotes the (de)convolution operation, and $\sigma_l$ the nonlinear operator (e.g., ReLU is used here). Here, $W_l$ corresponds to $n_l$ filters of support $n_{l-1}\times c\times c$, where $n_{l-1}$ is the number of channels in the previous layer, $c$ is the spatial size of a filter.

**2.2 Network training**

In our training, we opted to use patch-wise images instead of the original images as input. The patch processing cropped the original images into 21×21 overlapping patches, and the overlapping rate is 66%. The reason we used patch-wise processing is explained as follows. During network training, the memory of a single batch is limited. If we directly feed the original images into the training, the network can only take limited images in one batch because of memory overflow. Alternatively, if we divide the image into many overlapping patches and shuffle them, we will be able to provide abundant training data in a single batch and prevent overfitting, as seen and proven useful in many deep learning studies.[35,39,40] By using overlapping patches, we generated approximately 16,000,000 training samples from 40 HCP datasets.

During training, both the input (DWIs) and output (ground truth FA, MD or Eigenvectors) are given. The information is then used to train the filter weights and biases. Specifically, the objective is to minimize the loss function defined as the average mean squared error ($L_2$) between the network prediction of input $x^t$ and the corresponding ground truth result $y^t$ for all training samples $t = 1, \ldots, n$:

$$L(\Theta) = \frac{1}{n}\sum_{t=1}^{n}\|F(x^t;\Theta) - y^t\|^2, \quad (3)$$

where $F(\cdot)$ represents the operation performed by the neural network, and $\Theta$ denotes the filter weights $W_l$ and biases $B_l$. For multi-channel tasks (eigenvectors), the cost function is the averaged loss of all channels. The loss is minimized by adaptive moment estimation (ADAM), a gradient-based stochastic optimization algorithm implemented in Caffe.[43,44] The algorithm optimizes the loss by iteratively going through the network forward and backward and updating the network parameters with adaptive gradients and root mean square propagation. The loss computed with the training data pairs $(x^t, y^t)$ is known as the training loss. The validation loss is obtained by evaluating the updated network with the cross-validation data $(x^v, y^v)$.

## 3. METHODS

### 3.1 Data acquisition

**Human connectome project data**

DWI data from a total of 50 subjects were randomly selected from the Human Connectome Project (HCP Young Adult).[45] The diffusion-weighted scans were performed using an HCP-specific variant of the multiband diffusion sequence. The diffusion MRI data were collected with 3 different gradient tables, each including 90 diffusion weighting directions, plus six $b$ = 0 acquisitions. The diffusion directions were uniformly distributed on multiple q-space shells. The directions were optimized so that every subset of the first $M$ directions is also isotropic. Each dataset includes 18 non-DWIs and 270 DWIs in three different b values: 1000, 2000, and 3000 s/mm$^2$ and 90 diffusion directions. Data from 40 subjects (145 slices for each subject, a total of 5,800 images) were used for training and cross-validation (ratio of 4:1), and the data from the rest 10 subjects were used for testing and statistical analysis.

**Healthy volunteer and stroke patient data**

The data used in this study were acquired previously in an ongoing trial (Identifier: NCT03163758) as described in Li et al and Guan et al.[46,47] The study was approved by the ethics committee of the Peking Union Medical College Hospital. We selected data from 10 healthy volunteers (5 males and 5 females, age 54.4+/-7.8 years) and 2 stroke patients (30 and 76 years old). MRI data, including DWIs and structural images, were acquired on a 3.0T MRI scanner (MAGNETOM Skyra; Siemens, Erlangen, Germany) using a 20-channel phased-array head coil. The DWI images were acquired with a diffusion-weighted echo-planar imaging sequence (EPI) sequence. The DWI scan consisted of 30 diffusion-weighted directions with a b-value of 1000 s/mm$^2$ and one volume without diffusion weighting (i.e., b0 image). The parameters of the DTI sequence were as follows: repetition time (TR) = 7900 ms, echo time (TE) = 94 ms, slice thickness = 2.5 mm, field of view (FOV) = 240 mm × 240 mm, 60 axial slices with a slice thickness of 2.5 mm, slice gap 0.5 mm, matrix size = 122 × 122, and two repetitions. Earplugs and earphones were used to reduce scan noises, and the head motion was minimized by stabilizing the head with cushions. Based on the acquisition protocol of this data, the selected M= 6, 8, or 10 diffusion directions were uniformly distributed on the q-space shell with b=1000 s/mm$^2$. Data from 10 healthy volunteers (a total of 600 images) were used for training and cross-validation (ratio of 4:1), and the data from the 2 stroke patients were used for testing and analysis.

### 3.2 Data processing

The model-fitting results from all 90 directions of b=1000 s/mm$^2$ were used as the reference for training of the deep learning network, calculation of statistical metrics, and comparison of performance. A subset of the DWIs were chosen to generate the desired maps using both the proposed and state-of-art methods. Based on the acquisition protocol of the HCP data, the first M = 6, 18, or 36 diffusion directions were selected such that they are uniformly distributed on the q-space shell with b = 1000 s/mm$^2$. Such a uniform selection of the diffusion-encoding scheme improves both the rotational invariance of measurement precision and the robustness to noise.[48] MRtrix[49] was used to perform the conventional model fitting (MF) using the diffusion tensor model. Conventional MF, MLP (similar to Golkov et al[25] but for DTI), and BM4D[50,51] denoising followed by MF were used for comparison. All testing images were segmented into 286 regions of interest (ROIs) using a multiple-atlas likelihood fusion algorithm.[52] The mean values and errors in each ROI were calculated. The performance of different methods in representative gray matter, major white matter, and subcortical white matter structures were examined.

For the 3D fiber tracking generation, the fiber streamlines were assigned by Continuous Tractography (FACT)[53,54] with the "brute-force" method,[55] which was conducted with a fractional anisotropy threshold of 0.2, and a principal eigenvector turning angle threshold of 40° between two connected pixels. This generation was practically performed using the software DtiStudio.[56] Exploiting the manually delineated ROIs, an ROI-based approach was applied to construct the tracts of interest, confining all tracked pixels within the brain (the brute-force approach). In our work, we specifically defined three representative ROIs within one middle slice. They are functionally important white matter tracts: corpus callosum (CC), isolation of corticospinal tract (CST), and superior longitudinal fasciculus (SLF).

Image normalization was performed by dividing the image intensities by the maximum intensity of all images from the same subject. Training data were augmented with image rotations with 90° increments (to avoid interpolation errors) and image flipping from left to right. For all experiments, the testing data were strictly excluded from the training process. The training was performed using 2 NVIDIA Quadro P6000 graphics processing units (GPUs) with 24-GB memory capacity each. The learning rate was set empirically, and the number of epochs was selected based on the cross-validation error. The learning rate (weight of the negative gradient) was first set to 0.0002, momentum (weight of the previous update) 0.9, weight decay 0.0001 for the first 80 epochs. Then the learning rate was set to 0.0001 for the last 20 epochs to fine-tune the network parameters. The training process took a total of 10 hours. The convergence curves for training FA, MD, and colormaps of the HCP data is shown in Supporting Information Figure S1. The training and validation losses defined in Eq. (3) decreases with more epochs. The flat validation loss indicates a stable training regime with a good bias-variance tradeoff. The network parameters that correspond to the lowest validation loss were saved and used for the final testing.

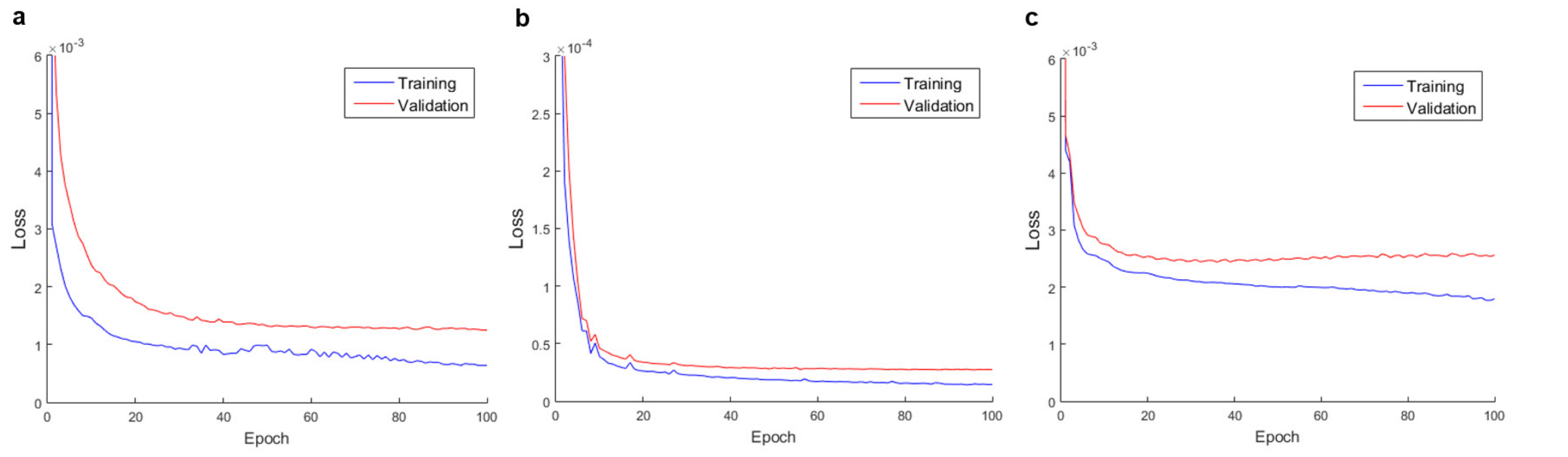

**FIGURE S1** Convergence curve of training and validation loss with respect to training epoch for (a) FA map, (b) MD map, and (c) colormap.

### 3.3 Evaluation metrics

Peak signal-to-noise ratio (PSNR), normalized mean squared error (NMSE), and structural similarity index (SSIM)[57] were used to quantify the similarity between the results of different methods compared to the reference. The lesion contrast was calculated as the FA value difference between the lesion and the surrounding background normalized by the mean FA value of the background. Large values suggest better contrasts.

## 4. RESULTS

### 4.1 FA and MD maps

The network in Figure 1 was trained (on 40 subjects) and tested (on 10 other subjects, i.e., $n$ = 10) using the diffusion data from the HCP dataset. Figure 2 shows representative FA maps from 6 DWIs generated using the conventional tensor model fitting (MF), multi-layer perceptron (MLP) (similar to Golkov et al[25]), block-matching and 4D filtering (BM4D) denoising algorithm,[50,51] and the proposed deep learning method (SuperDTI). Figure 3 plots the average mean values and the NMSEs for FA at several ROIs in gray matter structures, subcortical, and major white matter structures for all 10 testing datasets. Figure 4 shows representative MD maps from 3 DWIs using different methods. Results from MF using all 90 DWIs plus 18 non-weighted images were used as the ground truth. While other results became noisy or blurry in such an extreme case, the FA maps generated by our proposed method showed no apparent degradation. The generated FA/MD maps were quantitatively evaluated using the PSNRs, SSIMs, and NMSEs, which show the superior performance of SuperDTI even with only 6 DWIs for FA and 3 DWIs for MD maps. It is seen that the difference between the MD maps is far less apparent than that between the FA maps. This is because the MD calculation is less sensitive to noise.

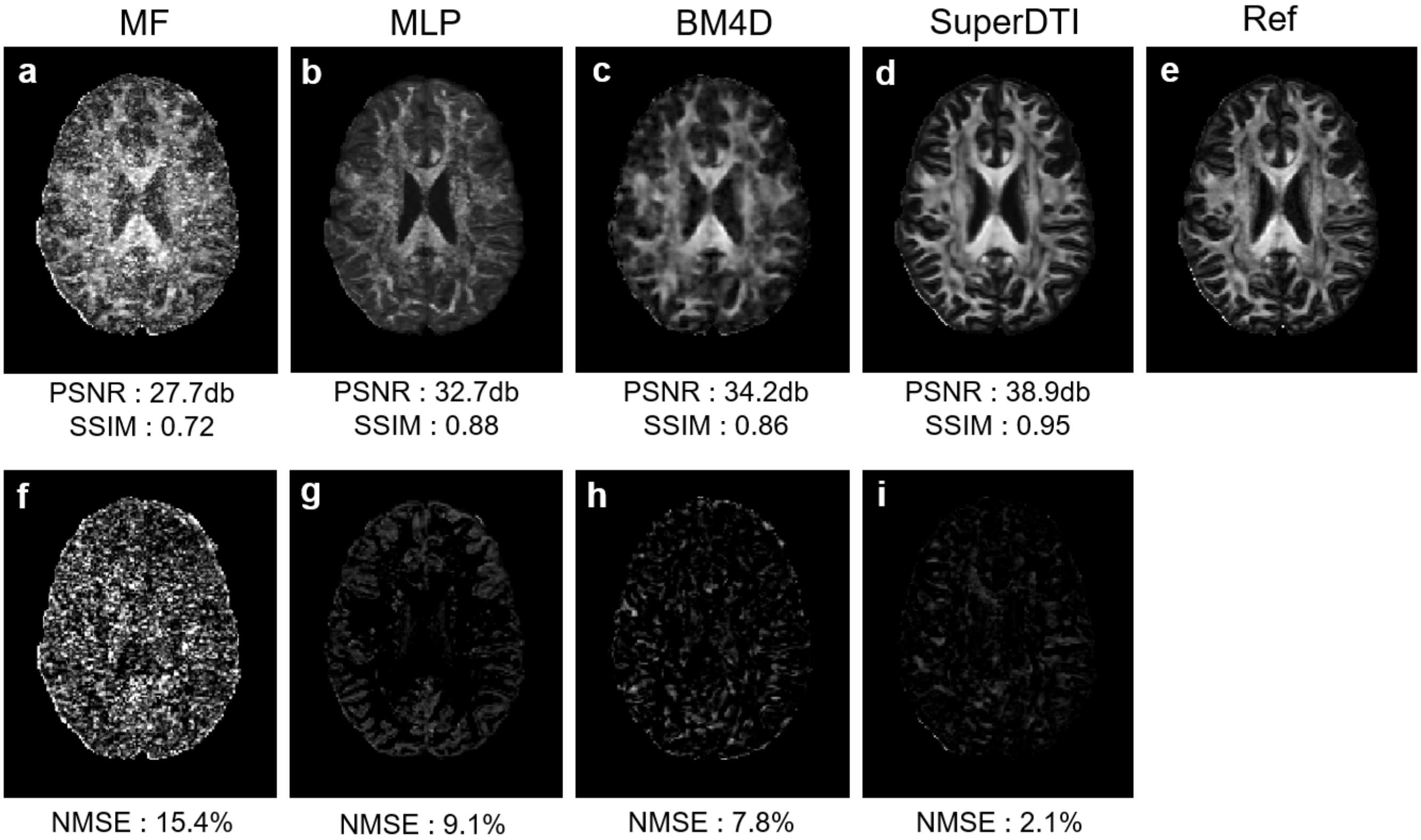

**FIGURE 2** Comparison of FA maps generated from 6 DWIs using different methods. FA maps generated by (a) MF, (b) MLP, (c) BM4D, (d) proposed SuperDTI, (e) reference and (f)-(i) the corresponding error maps respectively. The PSNRs, SSIMs, NMSEs, and error maps were calculated with (e) the model fitted FA map from 90 DWIs as the reference.

Within ROIs defined in the cortical gray matters, subcortical, and major white matter structures, the FA values estimated from limited DWIs using the proposed method also showed significantly less deviation from the ground truth than other methods. When only 6 DWIs were available, the conventional MF method significantly overestimated the FA values. In contrast, the SuperDTI had NMSEs below 3.5% for GM structures, below 1.5% for subcortical WM structures, and below 4% for major WM structures.

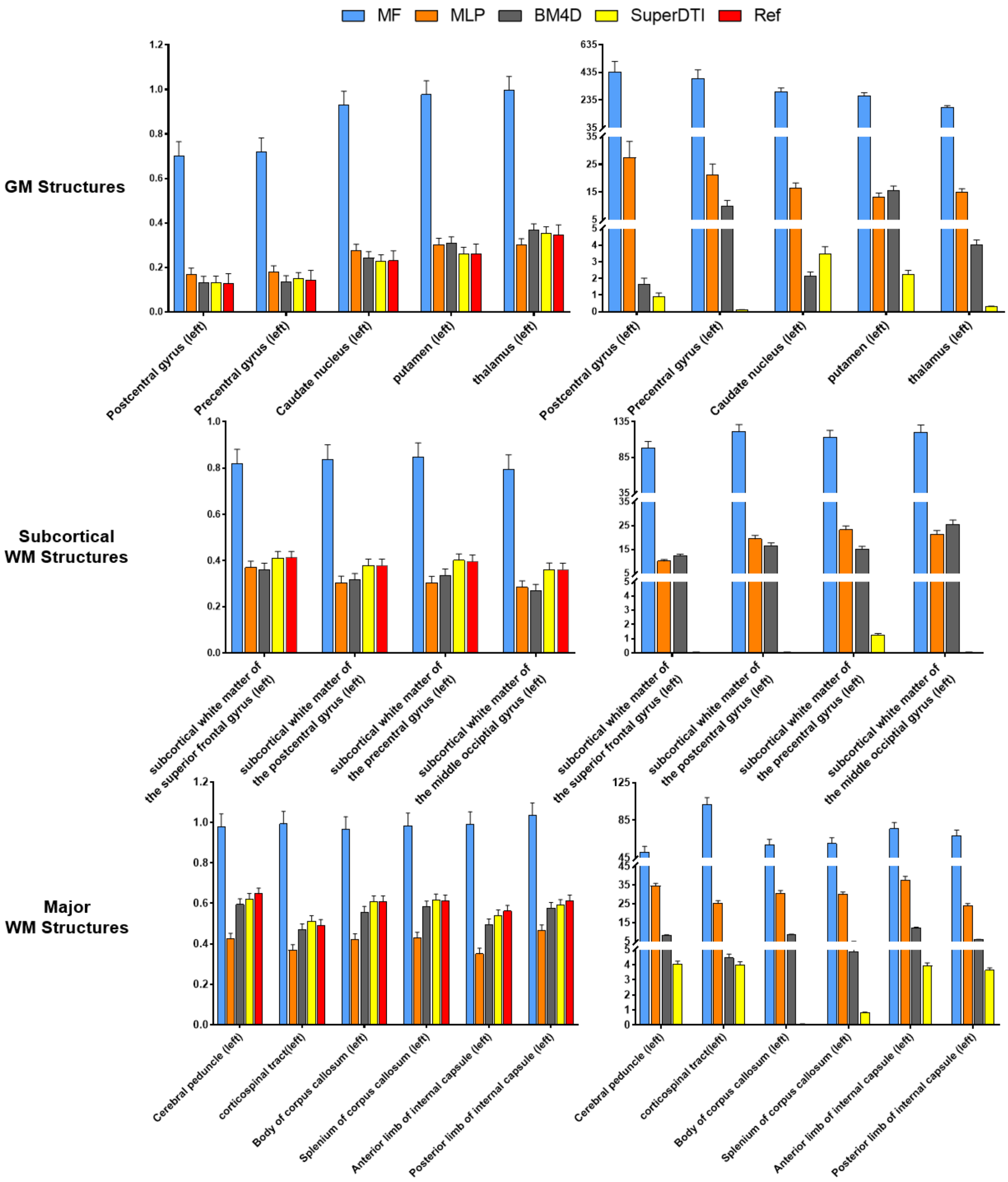

**FIGURE 3** Comparison of mean FA values (left) and the corresponding percentage error (right). Values and errors were obtained by MF, MLP, BM4D, and SuperDTI with 6 DWIs for different ROIs.

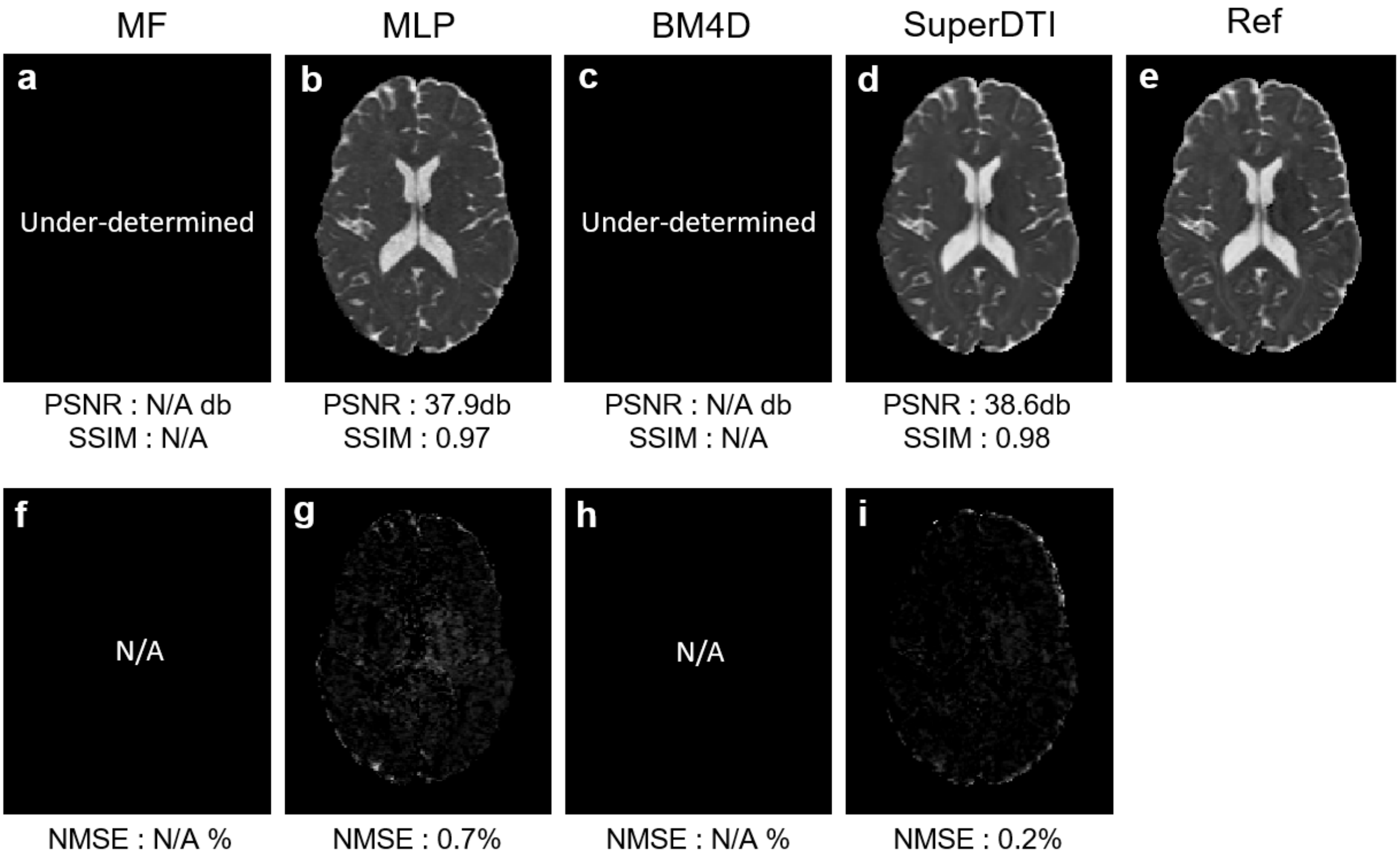

**FIGURE 4** Comparison of MD maps generated from 3 DWIs using different methods. MD maps generated by (a) MF, (b) MLP, (c) BM4D, (d) proposed SuperDTI, (e) reference and (f)-(i) the corresponding error maps respectively. The PSNRs, SSIMs, NMSEs, and error maps were calculated with (e) the model fitted MD map from 90 DWIs as the reference.

### 4.2 Directionally encoded colormap and fiber tracking

The primary eigenvector from a diffusion tensor indicates the principle orientation of water molecule diffusion. The primary eigenvector can be used to generate directionally encoded colormap, which is useful to inspect 3D orientation information encoded in diffusion tensors, as well as for fiber tracking. Our DL network can also be trained to estimate the primary eigenvectors directly from DWIs without MF. Figure 5 shows the directionally encoded colormap generated from the eigenvectors using the proposed and other competing methods with 6 DWIs. Visually, the directionally encoded colormaps generated using SuperDTI maintain the quality better than others with a small number of DWIs. Even with only 6 DWIs, the estimated orientation of the corpus callosum remains consistent along with the medial-lateral orientation[5] in SuperDTI, whereas other results show more speckles due to erroneous estimates. In Figure 6, fiber tracking results using the proposed method better preserve the morphology of three major white matter tracts in the brain than the others in the extreme 6 DWIs situation. Quantitatively, we used the number of fibers to evaluate the fiber tracking capability of different methods, as shown in Table 1. The proposed method generated 28608, 7480, and 6234 streamlines from CC, CST, and SLF, respectively, which are close to 28119, 8377, and 6297 from the reference data. The numbers from MLP, BM4D and MF are far fewer, indicating more miscalculated directions.

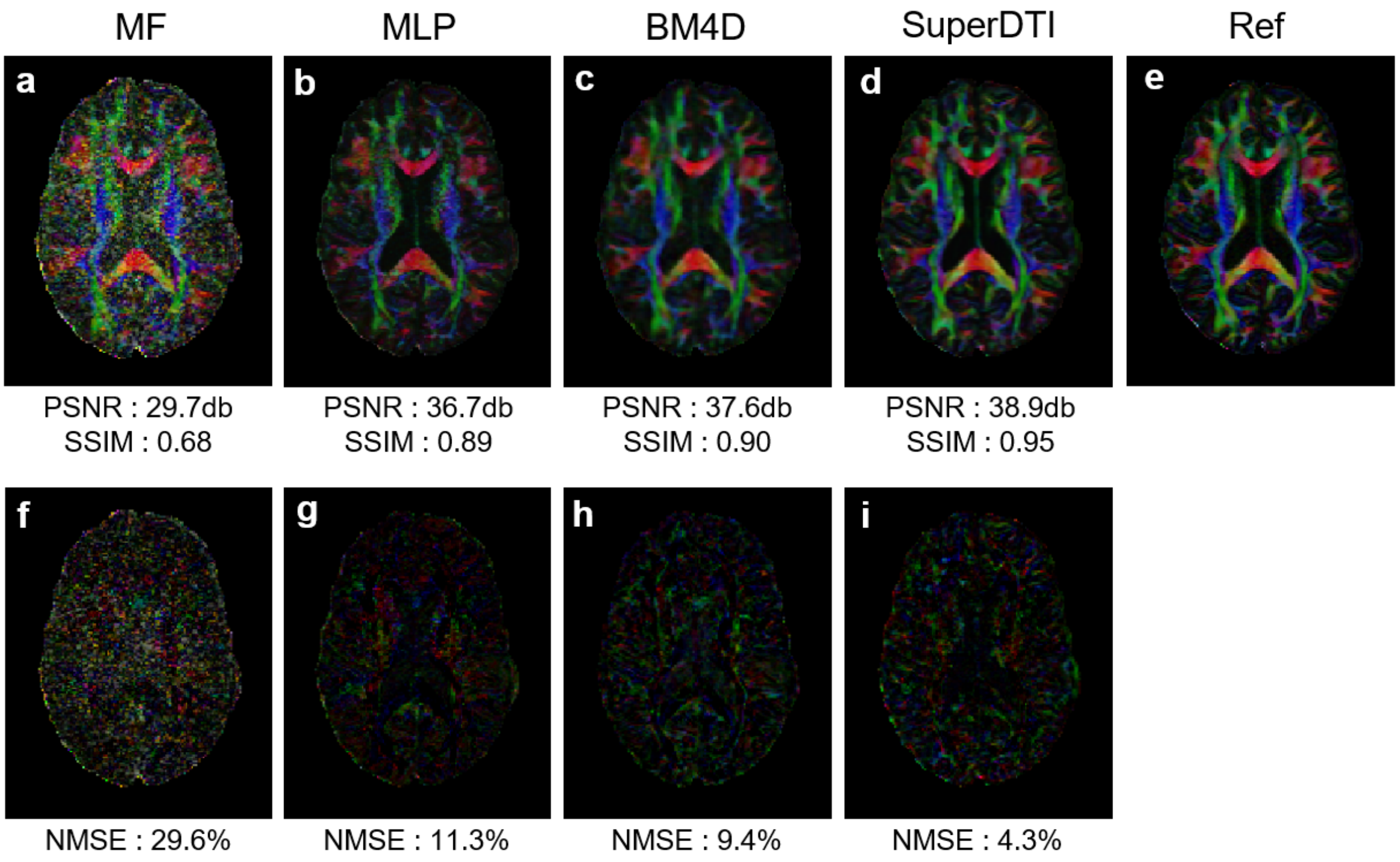

**FIGURE 5** Comparison of color-coded FA maps generated from 6 DWIs using different methods. The color maps generated by (a) MF, (b) MLP, (c) BM4D, (d) proposed SuperDTI, (e) reference and (f)-(i) the corresponding error maps respectively. The PSNRs, SSIMs, NMSEs, and error maps were calculated with (e) the model fitted colormap from 90 DWIs as the reference and shown with the average of all 3 color channels. The zoomed ROIs were circled in a red dotted line as shown in the middle row.

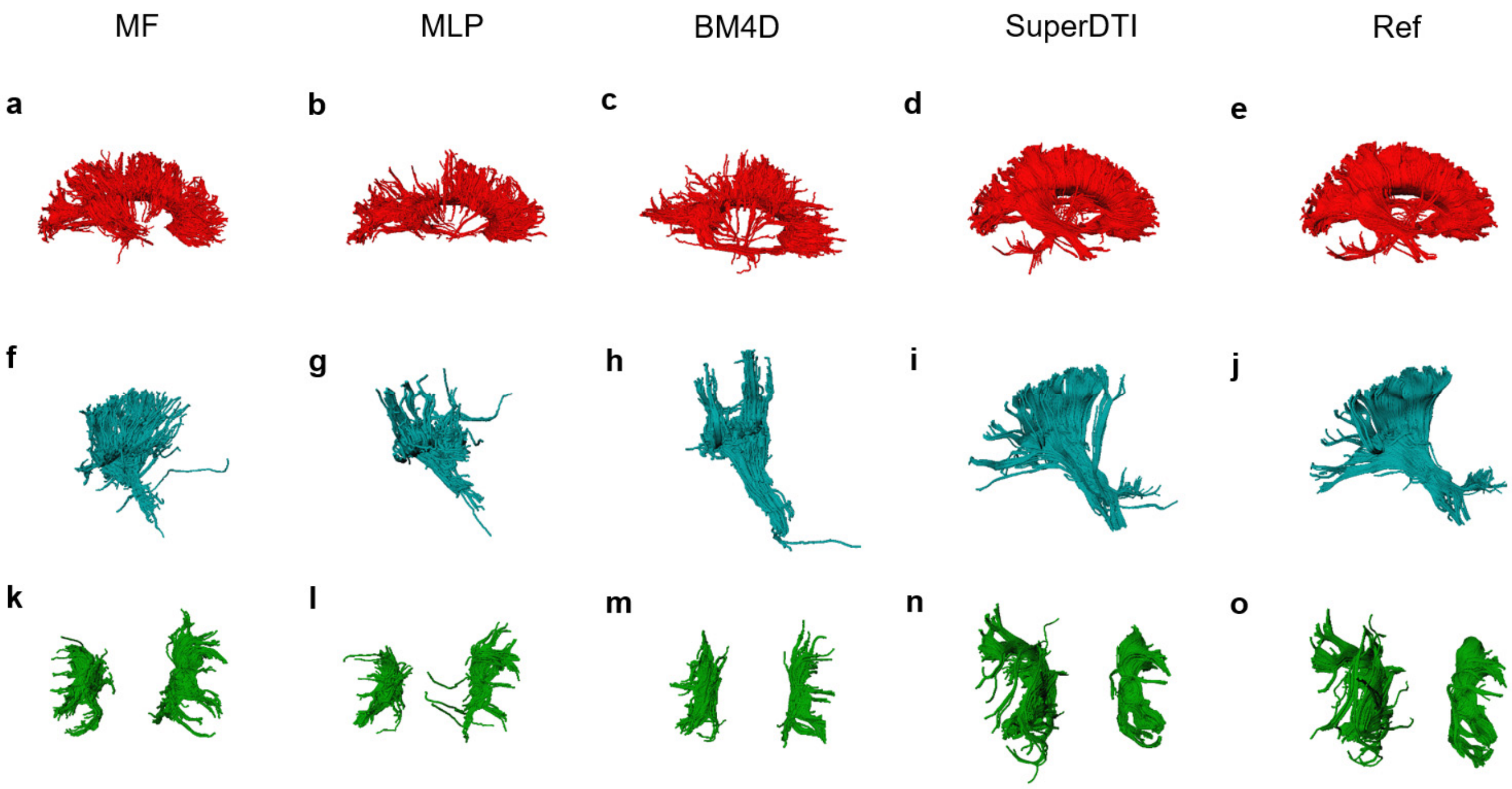

**FIGURE 6** Comparison of fiber tractography generated from 6 DWIs using different methods. Corpus callosum (CC), internal capsule/corticospinal tract (CST), and superior longitudinal fasciculus (SLF) generated by MF (a, f, k), MLP (b, g, l), BM4D (c, h, m), SuperDTI (d, i, n), respectively. Tractography from 90 DWIs (e, j, o) is shown as the reference.

**TABLE 1** Quantitative assessment of tractography using different methods

| **Methods** | | **MF** | **MLP** | **BM4D** | **SuperDTI** | **Ref** |
|---|---|---|---|---|---|---|
| **Fiber-number** | CC | 13911 | 16168 | 13706 | 28608 | 28119 |
| | CST | 3966 | 4882 | 2634 | 7480 | 8377 |
| | SLF | 3426 | 3737 | 2756 | 6234 | 6297 |

### 4.3 Statistical analyses

For the HCP data, ten sets of test images (a total of 1450 images) were used to evaluate the statistical accuracy of the FA, MD, and directionally encoded colormap values estimated using different methods. Quantitative assessments of PSNRs, NMSEs, and SSIMs averaged among all test images are summarized in Table 2.

**TABLE 2** Quantitative Assessment of FA, MD, and colormap using different methods

| Methods | | | MF | MLP | BM4D | SuperDTI |
|---|---|---|---|---|---|---|
| FA | 6 DWIs | PSNR | 27.15 | 32.89 | 33.96 | 38.60 |
| | | NMSE | 0.17 | 0.09 | 0.08 | 0.02 |
| | | SSIM | 0.69 | 0.88 | 0.85 | 0.95 |
| MD | 3 DWIs | PSNR | N/A | 37.97 | N/A | 38.61 |
| | | NMSE | N/A | 0.007 | N/A | 0.002 |
| | | SSIM | N/A | 0.97 | N/A | 0.98 |
| Color | 6 DWIs | PSNR | 30.12 | 36.95 | 37.93 | 41.34 |
| | | NMSE | 0.26 | 0.09 | 0.08 | 0.04 |
| | | SSIM | 0.71 | 0.90 | 0.91 | 0.95 |

### 4.4 Robustness to noise

In order to study the robustness to noise, we added additional 30dB Rician noise to the diffusion-weighted images. As shown in Supporting Information Figure S2, the result reveals that the conventional model-fitting method fails to generate acceptable FA maps, even using all 90 DWIs. The high sensitivity of MF[36] further demonstrates the benefit of bypassing the model fitting procedure. In contrast, our proposed method is still capable of generating accurate maps using as few as 6 noisy DWIs, regardless of whether the neural network is trained using noisy or clean DWIs.

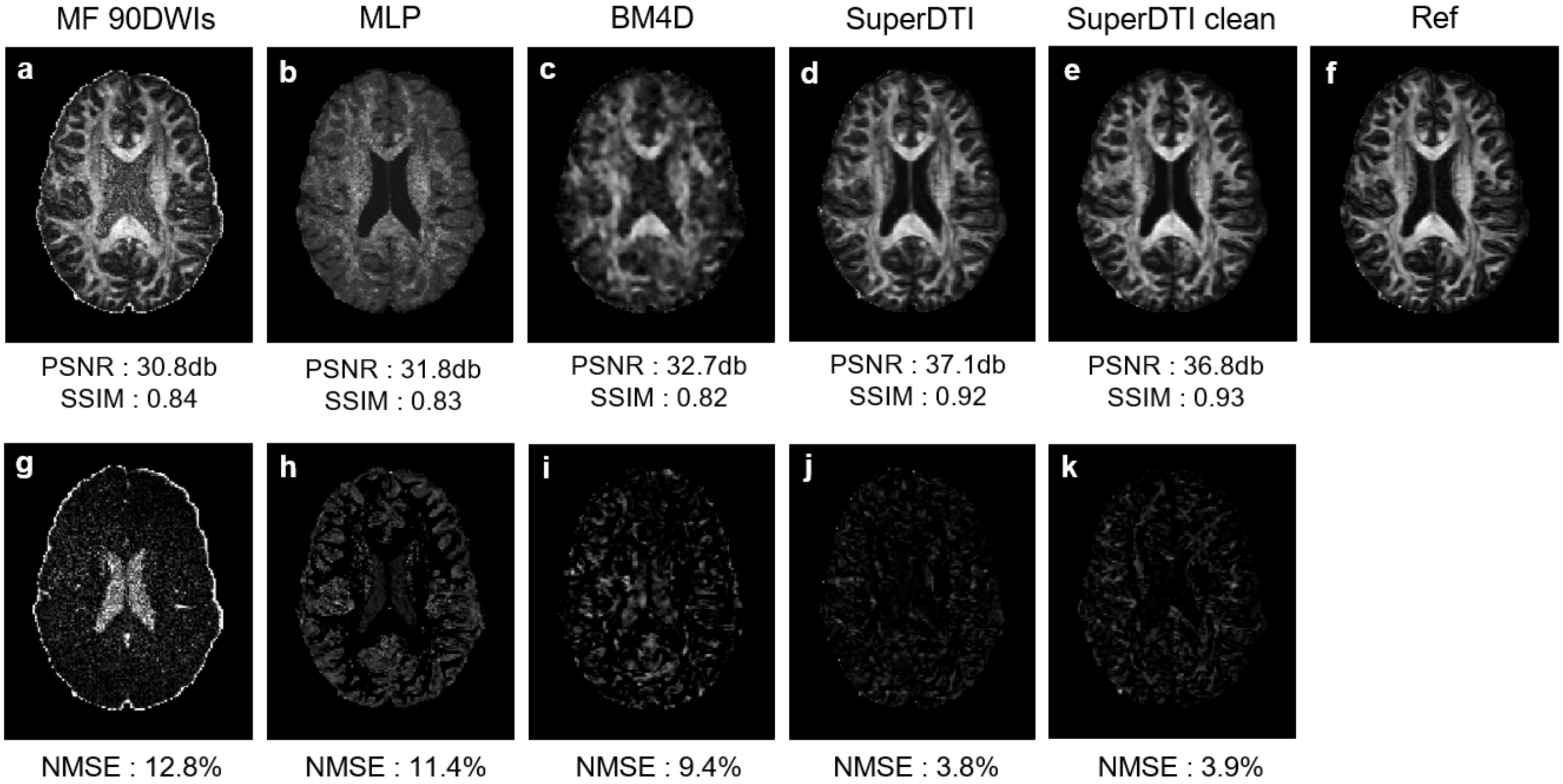

**FIGURE S2** Comparison of FA maps generated with noisy DWIs. (a)(f) FA maps generated using MF from 90 noisy and clean DWIs, respectively. FA maps generated from 6 noisy DWIs by (b) MLP, (c) BM4D, (d) proposed SuperDTI. Noisy DWIs were used for training in (b)(d). The SuperDTI clean (e) uses clean DWIs for training but 6 noisy DWIs as the network input. The PSNRs and SSIMs were calculated with the model-fitted FA map from 90 DWIs (f) as the reference.

### 4.5 Motion robustness

To evaluate the robustness of SuperDTI to motion misregistration from different DWI scans, we simulated motion by randomly selecting some DWIs and shifting the selected images by 1-3 pixels and/or rotating by 1 degree in a random direction. Such random selection, shifting, and rotation were repeated five times. SuperDTI trained without motion was tested on 6 DWIs with simulated motion. The qualitative and quantitative results of the reconstructed FA maps is shown in Supporting Information Figure S3. The PSNR and SSIM were the average values from five repetitions and 10 datasets. The model fitted FA maps from 90 DWIs were used as the reference. It can be seen that the FA map is still close to the reference even with simulated motion.

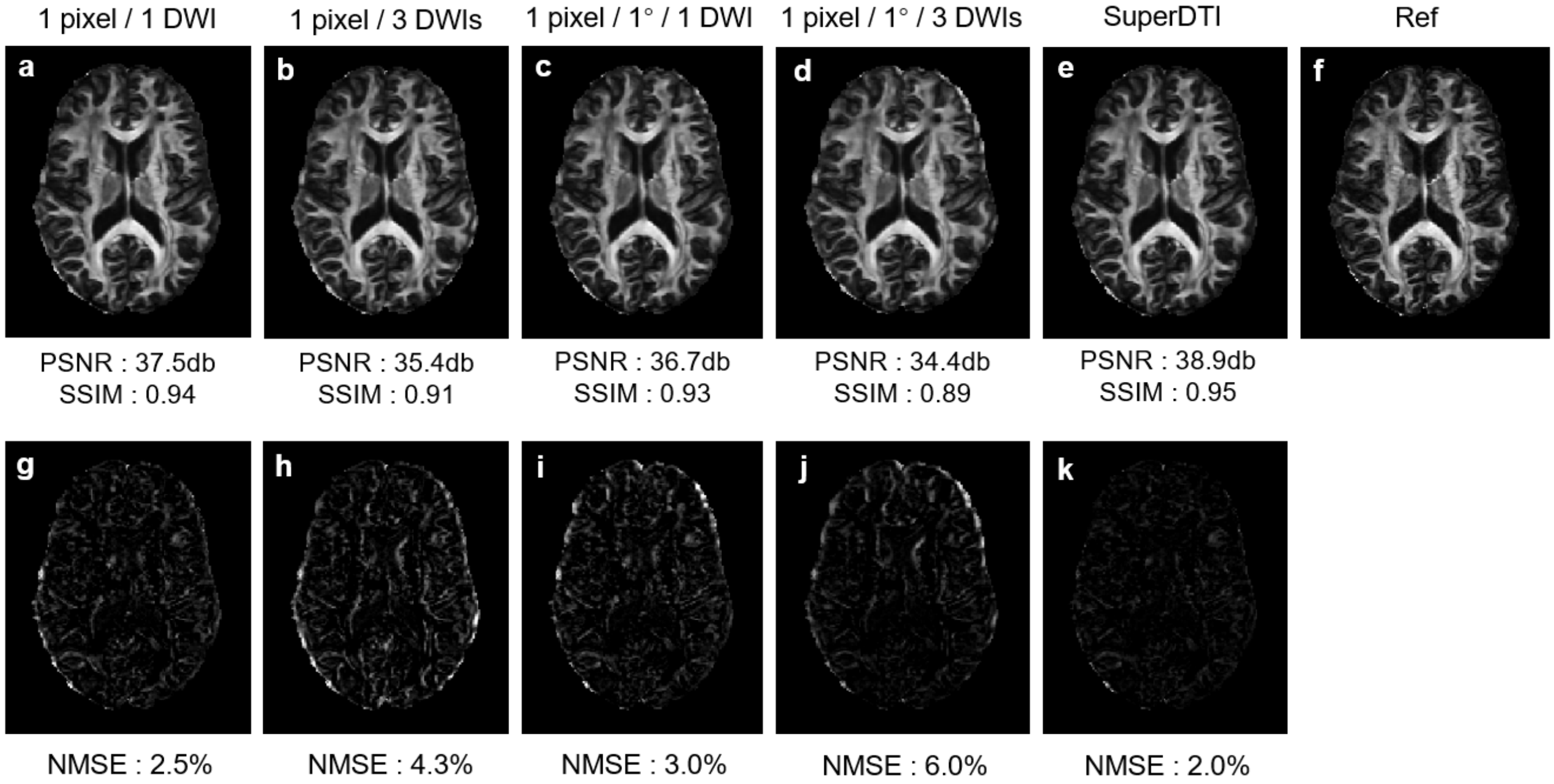

**FIGURE S3** SuperDTI generated FA maps from 6 DWIs with simulated motion. (a) shift of one pixel in one DWI; (b) shift of one pixel in 3 DWIs; (c) shift of one pixel and rotation of one degree in one DWI, (d) shift of one pixel and rotation of one degree in three DWIs; (e) no motion; and (g)-(k) the corresponding error maps respectively. (f) Reference.

### 4.6 Lesion detection in stroke patients

The proposed SuperDTI trained on 10 healthy volunteers was used to obtain the FA maps of two stroke patients. Figure 7 compares the FA maps obtained from 6, 8, and 10 DWIs using the proposed SuperDTI method with the FA map from 60 DWIs (30 directions with 2 averages) using the conventional tensor fitting as the reference. The relative contrast of the lesion is shown at the bottom of each image. It is seen that the FA map from 6 DWIs using SuperDTI can still reveal lesions with high contrast similar to that from 60 DWIs. Although the training was performed on healthy volunteer data, the network was still able to reveal the lesion information for stroke patients.

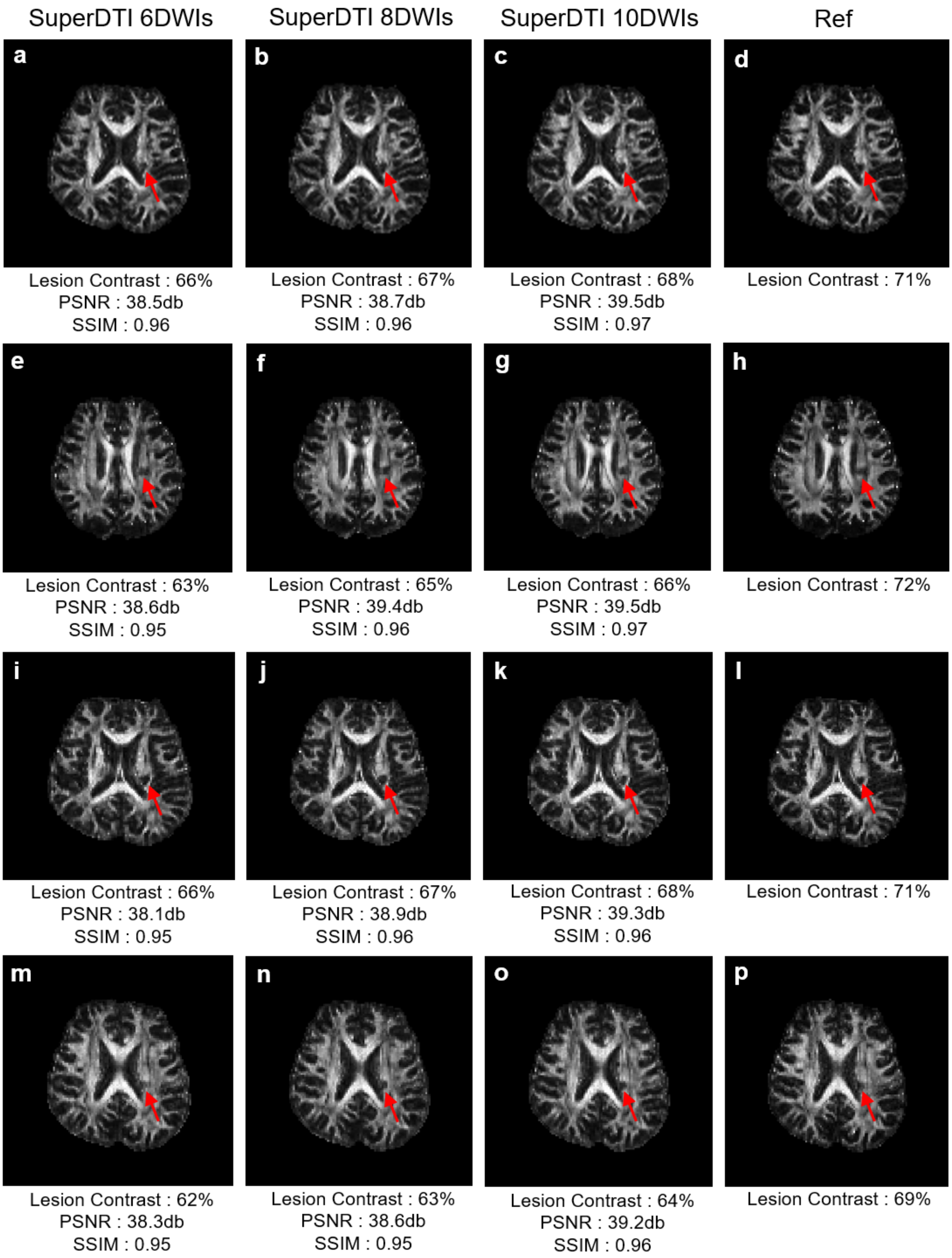

**FIGURE 7** Lesion detection using FA maps generated using SuperDTI. FA maps with a lesion (red arrow) generated by SuperDTI with 6 DWIs, 8 DWIs, and 10 DWIs (left to right columns) for different slices of two subjects (top to bottom rows). The PSNRs and SSIMs were calculated with the model-fitted FA map from 60 DWIs as the reference. The lesion contrasts were calculated for all images.

### 4.7 Effect of training size

**TABLE 3** Analysis of FA estimated by SuperDTI with different training sizes

| # datasets | | 40 sets | 20 sets | 10 sets | 8 sets | 6 sets | 4 sets | 2 sets |
|---|---|---|---|---|---|---|---|---|
| FA | PSNR | 39.3 | 38.9 | 39.2 | 38.6 | 39.0 | 38.2 | 36.7 |
| | SSIM | 0.95 | 0.95 | 0.95 | 0.94 | 0.95 | 0.94 | 0.90 |

We decreased the number of training datasets gradually and evaluated the accuracy of the FA values estimated by the proposed SuperDTI with 6 DWIs. The averaged NMSEs and PSNRs over all 10 testing datasets were summarized in Table 3. It can be seen that both NMSEs and PSNRs stay about the same as the training size decreases until only 2 datasets were used for training. The results demonstrate that by using overlapping patches for training, sufficient training data were used in our study without overfitting.

## 5 DISCUSSION

Although deep CNN has often been treated as a black box, it is important to understand how CNN works for a specific application. For example, the training size needs to be sufficiently large to avoid overfitting. For problems with global correspondence between the input and output (e.g., image classification where the output depends on all pixels of the input image), it is common to have tens of thousands of training datasets that contain various representations of such global correspondence. For problems with local correspondence (e.g., voxel-wise fitting of image intensities to a specific model[58]), significantly fewer training datasets are required. The results in Table 3 indicate that the proposed network does not need many training sets. This is because the values on the DWIs and those on the quantitative maps are local correspondence (theoretically pixel-wise correspondence). Each image already has a large number of pixels, which provide a large number of training sets. As a result, only a few images are needed for training the proposed network. Although it is always a challenge to predict the minimum training size required, the 5,800 training images used in this study were shown empirically to be sufficient to ensure robustness and avoid overfitting of the proposed network. We chose CNN over MLP because MLP has the characteristic of fully connected layers, where each node is connected with every node in the adjacent layer. As a result, the total number of trainable parameters in MLP can grow exponentially. In contrast, CNN is more much efficient by taking advantage of the spatial correlation in image pixels.

Another critical issue is whether the network trained using a set of data can be generalized to a separate set of data acquired on the same scanner with the same protocol. Our results in Figure 7 demonstrate that the proposed SuperDTI preserves the ability to detect lesions even though the network was trained using data from normal subjects. This ability can be explained by the local relationship between DWIs and FA maps. Since FA values for pathological tissues are relatively independent of the structure and content of the image (e.g., the shape and location of lesions), it is likely the outputs of a network trained to represent the local relationship can preserve the sensitivity of FA difference between normal and pathological tissues.

It is worth noting that the current version of the proposed method is not able to resolve fiber crossing because only tensor-derived parameters are estimated by the network. More DWIs from more diffusion directions are necessary to perform fiber tracking with the capability of resolving crossing fibers.[26] Since DTI-derived tractography has demonstrated clinical relevance,[59,60] this study is to enable superfast tractography in clinical settings.

We do not expect our model trained with one protocol/scanner to be directly applicable on another protocol/scanner due to the difference that cannot be characterized precisely, such as the difference in gradient table. However, for the same protocol and scanner, we have shown that the trained network is generalizable to new data from other subjects and with motion and different SNR. Such an ability to capture the tensor-derived metrics from 6 directions is still significant as the training only needs to perform once, and the trained network can be used for all future scans on the same scanner with the same DWI protocol. Characterization of generalizability to different protocols and scanners is beyond the scope of this paper and may be explored in our future studies.

## 6 CONCLUSION

In this paper, we demonstrated the feasibility of superfast diffusion tensor imaging and fiber tractography using SuperDTI with only six DWIs (that could be noisy and misregistered by motion). The proposed SuperDTI outperformed the conventional tensor model fitting and other state-of-the-art methods using both qualitative and quantitative metrics. We also demonstrated that the neural network trained on healthy volunteer datasets could be directly applied to stroke patient's data without compromising the lesion detectability. The network trained using clean, motion-free DWIs could be applied to noisy or motion-misregistered DWIs with high FA accuracy. In summary, the proposed SuperDTI is expected to benefit a wide range of clinical and neuroscientific studies that require superfast but reliable DTI.

## ACKNOWLEDGMENTS

This work is supported in part by the National Institute of Health Brain Initiative R01EB025133.

**SUPPORTING INFORMATION**

Additional Supporting Information may be found online in the Supporting Information section.